\documentclass[prl,twocolumn,reprint,10pt]{revtex4-1} 
\usepackage{amsmath,amssymb,graphicx}

\begin{document}

\title{Cavity QED on a nanofiber using a composite photonic crystal cavity}
\author{Ramachandrarao Yalla}
\thanks{These authors contributed equally to this work.}
\author{Mark Sadgrove}
\thanks{These authors contributed equally to this work.}
\author{Kali P. Nayak}
\author{Kohzo Hakuta}
\email{email address:hakuta@cpi.uec.ac.jp}
\affiliation{
Center for Photonic Innovations, The University of Electro-Communications, Chofu, Tokyo 182-8585, Japan
}
\begin{abstract}
We demonstrate cavity QED conditions in the Purcell regime for single quantum emitters on the surface of an optical nanofiber. The cavity is formed by combining
an optical nanofiber and a nanofabricated grating to create a composite photonic crystal cavity. Using this technique, significant enhancement of the spontaneous 
emission rate into the nanofiber guided modes is observed for single quantum dots. Our results pave the way for enhanced on-fiber light-matter interfaces with clear applications to quantum networks.
\end{abstract}
\maketitle
\maketitle
Cavity based enhancement of light-matter interactions - referred to as cavity quantum electro-dynamics (QED) - represents a major
advance in our ability to control single quantum emitters (QEs) and single photons. One motivation in this field is
the possibility of using QEs coupled to cavities as nodes in a quantum network~\cite{Kimble}. 
Recently, nanophotonic cavity QED devices have attracted great interest~\cite{PhCReview,NodaReview},
with numerous studies achieving the Purcell regime~\cite{Purcell} of cavity QED for QEs
in effective 1D photonic crystal (PhC) cavity structures~\cite{MangaRao,LundHansen,Englund,Hung,Thompson}.

Among these PhC cavity devices, various types of nanowaveguide cavities have proved to be promising 
with recent implementations including diamond nanobeams~\cite{Hausmann}, silicon nitride alligator waveguides~\cite{Goban} 
and PhC nanofiber cavities~\cite{KaliFIB,KaliPhC2}. 
However, for application to quantum networks, in-line (i.e. fiber integrated) light-matter interfaces such as those 
realized by optical nanofibers are advantageous
since automatic coupling to a single mode fiber is achieved~\cite{Klimov,LeKienBF,Chandra,RauschenbeutelNanoTrap,KimbleNanoTrap}. 
Although direct fabrication of PhC cavities on nanofibers has seen recent progress~\cite{KaliPhC2,KaliPhC1,KaliFIB},
designability of the PhC parameters is still limited using this technique. 

Here we demonstrate a unique method to achieve cavity QED based enhancement of spontaneous emission (SE) from a single QE
 on an optical nanofiber. Our method is to create a composite PhC cavity 
(CPCC) by bringing a nanofiber and a nanofabricated grating 
with a designed defect into optical contact. 
\begin{figure}
\centering
\includegraphics[width=1\linewidth]{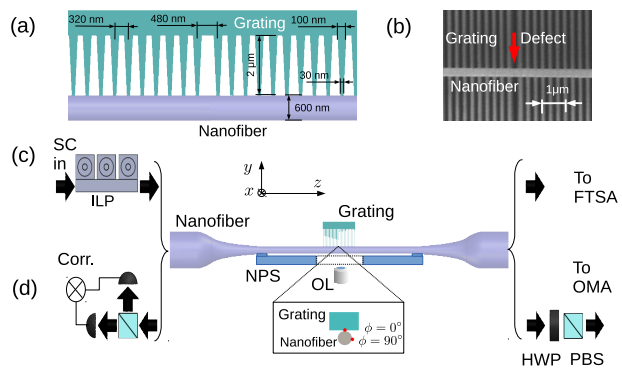}
\caption{\label{fig:device} (Color online) (a) Conceptual diagram of the device and design parameters.
(b) Scanning electron microscope image of the device. 
(c) Experimental setup used to characterize the optical response of the CPCC. 
 SC, ILP, and FTSA denote the super-continuum source, in-line polarizer, and Fourier transform spectrum analyzer respectively.
 (d) Experimental setup for PL intensity spectrum measurements. 
 The inset depicts the azimuthal position defined by angle $\phi$ of the QDs (red dots).
 Corr., NPS, OL, HWP, PBS, and OMA denote the correlator, nanopositioning stage, objective lens, half wave plate, polarizing beam splitter, and optical multi-channel analyzer respectively. 
 The $x-$, $y-$ and $z-$ axes are defined as shown.}
\end{figure}
The nanostructured grating as depicted in Fig.~\ref{fig:device}(a) was designed for an operating wavelength around $800$ nm. The grating was fabricated 
on a silica substrate using electron beam lithography along with chemical etching to create a grating pattern with trapezoidal slats extending $d = 2\;\mu$m from the substrate.
The period of the grating is $\Lambda_g=320$ nm and the slats have a tip width of $\alpha_t\Lambda_g=30$ nm and a base width of $\alpha_b\Lambda_g=100$ nm where $\alpha_t$ and $\alpha_b$ 
are the grating duty cycles at the tip and base of the slat respectively.
In the center of the grating pattern, a defect of width $3\Lambda_g/2 = 480$ nm was opened between the slats on either side. The number of slats was $N=350$.
The diameter $2a$ of the nanofiber at the point where the grating was mounted was between 550 nm and 600 nm. 
Figure~\ref{fig:device}(b) shows a scanning electron microscope image of the device where the nanofiber can be seen mounted on the grating
and crossing the defect region. 

Figures~\ref{fig:device}(c) and \ref{fig:device}(d) show the experimental setup for the optical characterization of the CPCC and the measurement of photoluminescence (PL) intensity spectra respectively.
In Fig.~\ref{fig:device}(c), a linearly polarized super-continuum source with an output wavelength range 
spanning from 700 nm to 1000 nm was introduced to the CPCC via an in-line polarizer. The resulting output spectrum was
measured using a Fourier transform spectrum analyser with a resolution of 0.01 nm. 
We deposited colloidal
quantum dots (QDs) with a nominal emission wavelength of 800 nm on the nanofiber by lightly 
touching the top surface of the nanofiber with a droplet of QD solution~\cite{ChandraOE}.
The QDs were thus distributed with an
azimuthal position  expected to be randomly distributed between $\phi=0^\circ$ and $\phi=90^\circ$~\cite{Chandra}. The definition
of $\phi$ is shown in the inset of Fig.~\ref{fig:device}(d).
The optical loss per deposition was estimated to be 0.6$\%$. 
We estimated the number of QDs by observing the blinking statistics at each deposition. To confirm single QDs,  we
performed photon correlation measurements using a Hanbury-Brown-Twiss setup as 
depicted in Fig.~\ref{fig:device}(d). (Details may be found in Ref.~\cite{ChandraOE}). The QDs were excited using a 640 nm wavelength laser of power $35\;\mu$W focused
by an objective lens~\cite{ChandraOE}.
\begin{figure}
\centering
\includegraphics[width=1\linewidth]{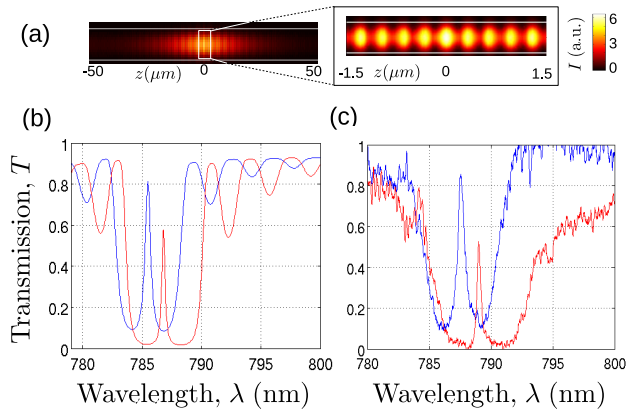}
\caption{\label{fig:cavity}(Color online)  (a)  Simulated electric field intensity inside the CPCC for the $y-$mode. The inset shows an expanded view of
the central region of the cavity. White horizontal lines mark the position of the edges of the nanofiber.
(b) and (c) Simulated and measured transmission spectra  respectively for the $x-$ (blue line) and $y-$mode (red line).}
\end{figure}
Using a nano-positioning stage, we positioned the QD in the center
of the excitation spot to better than $\pm1\;\mu$m by monitoring the QD fluorescence.
Through the objective lens, the defect was visible as a line in the center of the grating pattern.
We aligned the defect position to within $\pm1.5\;\mu$m of the excitation spot center.

To estimate the enhancement of SE, we measured the PL intensity spectrum using an optical multi-channel analyzer (OMA) (Fig. \ref{fig:device}(d)). 
The frequency domain response function of the OMA was measured using a single frequency laser, giving resolutions of $\Delta\lambda_{\rm OMA}^{1800}=0.1$ nm,
$\Delta\lambda_{\rm OMA}^{300} = 1.5$ nm, and $\Delta\lambda_{\rm OMA}^{150} = 3.3$ nm, where the superscripts indicate the lines/mm of the OMA gratings.
To resolve the polarization dependence of the cavity resonance, we used an OMA resolution of $0.1$ nm.
For single QD measurements, we used resolutions of $1.5$ nm and $3.3$ nm to increase the 
signal-to-noise ratio and to allow accurate determination of the background QD PL intensity.
It should be noted that the measured enhancement lineshape is given by
$E_{\rm meas.}(\lambda) = E_{\rm true}(\lambda) * h_{\rm OMA}(\lambda)$, where  $E_{\rm true}$ is the true enhancement spectrum, 
$h_{\rm OMA}$ is the OMA response function and $*$ denotes convolution. 
When $\Delta\lambda_{\rm true}\ll\Delta\lambda_{\rm OMA}$,
$\Delta\lambda_{\rm meas.}\approx\Delta\lambda_{\rm OMA}$ 
and the height of the measured enhancement peak
is reduced by the factor $\Delta\lambda_{\rm true}/\Delta\lambda_{\rm OMA}$.

In Fig.~\ref{fig:cavity}(a), we show the finite-difference time-domain simulated electric field intensity at the top surface of the nanofiber
at a resonance wavelength of $\lambda_{\rm res}=785$ nm for the
$y-$polarized cavity mode ($y-$mode). 
The electric field intensity of the cavity mode reduces exponentially by a factor of $1/e$ over a distance of $z_0 = 14\;\mu$m 
from the cavity center. We take the value $L_{\rm eff}=2z_0=28\;\mu$m as the effective length of the cavity. 
The inset shows the cavity mode over a region $\pm1.5\;\mu$m about the cavity center where the QD is expected to be positioned.
It may be seen that the peaks of the cavity mode over this region  are approximately of the same intensity with a variation of $\pm10\%$.
This implies that alignment with the exact cavity center is
not critical, as any one of the cavity antinodes in this region will lead to approximately the same enhancement of SE.

Due to the asymmetrical index modulation 
induced by the grating, the degeneracy of the $x-$ and $y-$ polarized fundamental modes ($x$- and $y$-modes) of the nanofiber is lifted.
Figures~\ref{fig:cavity}(b) and~\ref{fig:cavity}(c) show simulated and measured cavity transmissions
for the $x-$ and $y-$modes (blue and red lines respectively). 
For the simulations, we set $2a=550$ nm.
The $x-$ and $y-$mode resonance peaks are separated by 1.3 nm (simulations) and 1.4 nm (experiments).
The $x-$ and $y-$mode stop-band minimum transmission
values and the $x-$mode peak transmission value agree with the simulation values within the experimental error, 
while the experimental $y-$mode transmission is less than the simulation value by $16\%$.
The simulation (measured)  $x-$ and $y-$mode quality factors (Q-factors) were $1410$ ($1270\pm20$) and $2590$ ($2310\pm80$).
We note that the experimental and simulation results both clearly show that the $y$-mode has a larger 
Q-factor than the $x$-mode. This is because the $y$-mode experiences more modulation due to the grating leading to larger reflectivity of the Bragg mirrors and thus a higher Q-factor. 
The experimentally measured $x-$ and $y-$mode resonance peaks have Q-factors about 10$\%$ lower than the values predicted by simulations. 
We also measured the Q-factors
for $x-$ and $y-$mode resonance peaks over a range of wavelengths from 780 nm ($2a\sim550$ nm) to 800 nm ($2a\sim610$ nm). The Q-factor was found to increase as the wavelength became shorter as we describe later.
Additionally, we note that the exact value of $\lambda_{\rm res}$ for both the $x-$ and $y-$ modes is dependent on $a$~\cite{Mark}. By systematically measuring $\lambda_{\rm res}$ at 
different $a$, the rate of change was found to be 
$\Delta\lambda_{\rm res}/\Delta a = 0.60\pm0.03$.
\begin{figure}
\centering
\includegraphics[width=1\linewidth]{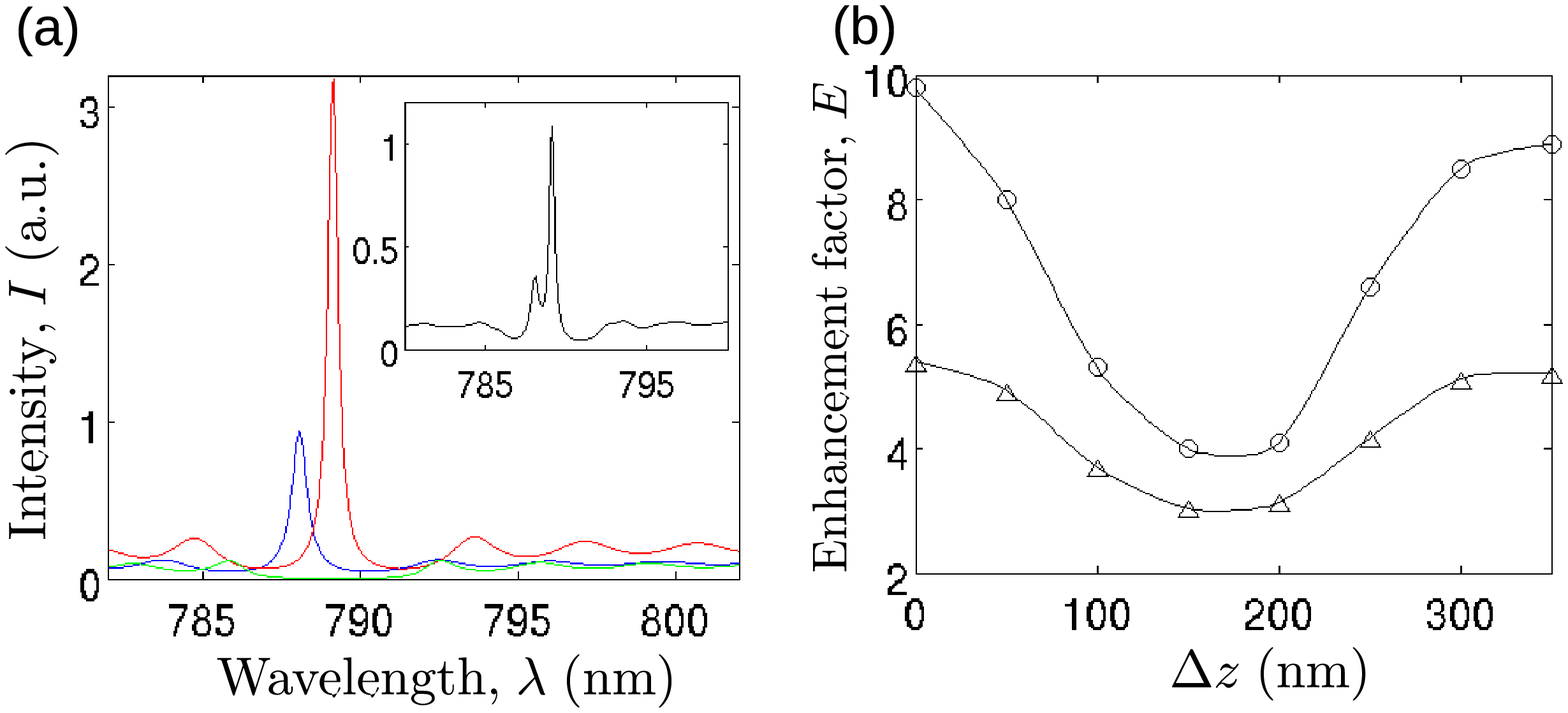}
\caption{\label{fig:simenhance}(Color online) (a)  Simulations of PL intensity spectra for $x-$ (blue line), $y-$ (red line), and $z-$ (green line) polarized dipole sources. 
The inset shows the average of the PL intensity spectra. 
(b) The dependence of the EF on the position $\Delta z$ of the dipole source in the cavity. Circles show results for $\phi=0^\circ$ and
triangles show results for $\phi=90^\circ$. 
}
\end{figure}

In Fig.~\ref{fig:simenhance}(a), we show simulated PL intensity spectra through the nanofiber for $x-$, $y-$, and $z-$ 
polarized dipole sources placed at the center of the CPCC for $2a=570$ nm. 
The lack of coupling for the $z-$ polarized dipole source is caused by the relative phase shift of the 
$z-$ component of the cavity mode~\cite{LeKien}. 
Enhancement factors (EFs) were calculated from the ratio of on- and off-resonance PL intensity spectra.
The peak EFs were found to be $10$ and $19$ for $x-$ and $y-$ polarized dipole sources respectively. 
The $x-$mode ($y-$mode) enhancement peak has $\lambda_{\rm res}^x=788.0$ nm ($\lambda_{\rm res}^y=789.1$ nm) and the $x-$mode ($y-$mode) full-width at half-maximum (FWHM) was 0.6 nm (0.4 nm).
To simulate the polarization averaged PL intensity spectrum, we averaged the results of Fig.~\ref{fig:simenhance}(a) (see inset).
The average EF was found to be 9.8 at the $y-$mode resonance wavelength. The drop in the EF relative to that for a $y-$polarized
dipole source is due to the increase in the off-resonance background PL intensity due to the contributions of the $x-$ and $z-$ polarized dipole sources.
By referencing the background PL intensity away from resonance to 0, the ratio of the $y-$ to $x-$mode peak values was calculated to be 3.0.

Figure~\ref{fig:simenhance}(b) shows simulations of the 
polarization averaged EF as a function of the displacement $\Delta z$ from the cavity center for dipole source azimuthal positions with angles 
of $0^\circ$ (circles) and $90^\circ$ (triangles).
The solid lines are interpolations to guide the eye. 
Note that the EF varies from a maximum of 9.8 for $\phi=0^\circ$ at the cavity center to a minimum of 3 for $\phi=90^\circ$ at the first cavity node.

\begin{figure} 
\centering
\includegraphics[width=1\linewidth]{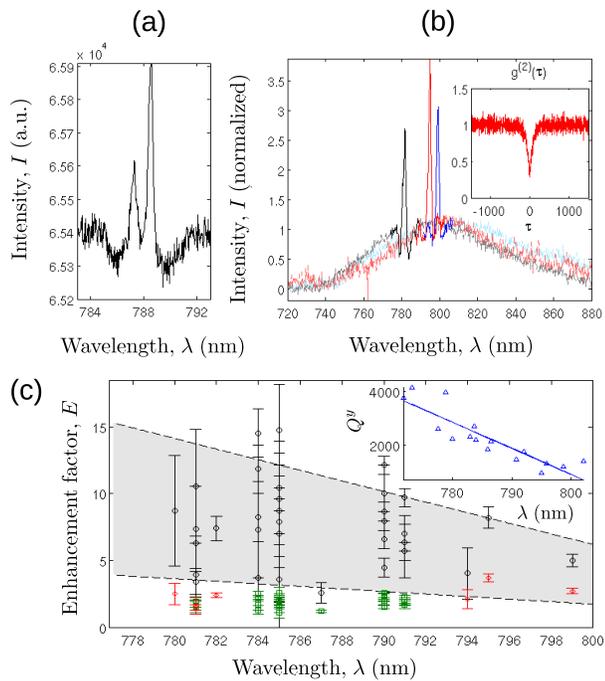}
\caption{\label{fig:enhance}(Color online) (a) PL intensity spectrum showing the $x-$ and $y-$modes (left and right peak respectively). 
(b) Measured PL intensity spectra for three different single QDs. 
The inset shows a typical normalized photon correlation signal.
(c) EFs for single QDs in the CPCC and cavity Q-factors. 
Red diamonds and green squares with error bars show measured EFs,
while black circles show corrected EFs. 
The inset shows experimentally measured Q-factors for the y-mode cavity transmission
peaks (linear fit shown by solid line).
}
\end{figure}

Figure~\ref{fig:enhance}(a) shows a typical measured polarization averaged PL intensity spectrum for an OMA resolution of $0.1$ nm 
and for a number of QDs between 3 and 5 as estimated from blinking statistics~\cite{ChandraOE}. 
Two enhancement peaks were clearly resolved with $\lambda_{\rm res}=787.3$ nm and $\lambda_{\rm res}=788.5$ nm 
assigned to the $x-$ and $y-$mode respectively based on independently performed 
polarization filtered measurements. 
The $x-$mode ($y-$mode) FWHM was $0.5\pm0.2$ nm ($0.4\pm0.1$ nm).
Because the spectrum of the QD ($50$ nm FWHM) is broader than the wavelength detection range 
of the OMA at the $0.1$ nm resolution setting ($30$ nm), the background PL intensity cannot be determined and therefore the 
EF cannot be calculated.
The ratio of the $y-$ and $x-$mode peak values was found to be $2.4\pm0.6$.
The experimentally determined ratio and the enhancement peak wavelength separation 
show good correspondence with the simulation values.
The good agreement between  simulations and experimental results suggests that the CPCC functions essentially as designed.

Figure~\ref{fig:enhance}(b) shows experimentally measured PL intensity spectra for three different single QDs at different positions on the nanofiber. 
The off-resonance PL intensity near $\lambda_{\rm res}$ was normalized to 1.
The inset shows a typical anti-bunching signal measured for the deposition where $\lambda_{\rm res}$ was 795 nm. 
The normalized intensity correlation function has a zero delay value of $g^{(2)}(\tau=0)=0.3$ indicating a single QD~\cite{ChandraOE}.
Sharp enhancement peaks can clearly be seen rising above the broad background PL intensity spectra of the QDs.
Although the $x-$ and $y-$ mode peaks are not resolved at this OMA resolution ($1.5$ nm), we assign the peak seen in 
the PL intensity spectra to the $y-$polarized dipole component since the $y-$mode peak is larger as seen in Fig.~\ref{fig:enhance}(a). 
The measured EFs were $2.7\pm0.2$, $3.9\pm0.3$, and $3.0\pm0.2$ at $\lambda_{\rm res}=782.0$ nm, $795.0$ nm, and $799.0$ nm 
(black, red and blue curves respectively). 

Figure~\ref{fig:enhance}(c) summarizes our results regarding the EF for single QDs along with measured Q-factors. The shaded region shows where the 
EF is expected to lie assuming random placement of the QD in the cavity and $\phi$ randomly distributed between 0$^\circ$ and 90$^\circ$.
The top and bottom black dashed lines, which are estimates of the maximum and minimum  simulated EFs respectively (see Fig.~\ref{fig:simenhance}(c)),
are found using a linear fit to simulated EFs at several wavelengths between $780$ nm and $800$ nm.
The simulated EF is seen to increase as $\lambda_{\rm res}$ becomes shorter. This may be explained by the
increase in the Q-factor as shown by the blue triangles in the inset of Fig.~\ref{fig:enhance}(c).

The red and green points in Fig.~\ref{fig:enhance}(c) show
the measured EF as a function of $\lambda_{\rm res}$ for $11$ different single QDs
with estimated error bars, for resolutions of $1.5$ nm and $3.3$ nm respectively. 
Different points at the same wavelength show data taken for separate grating mounting events for the same
QD. The variation in the EF may be understood as being due to the variation in the relative position 
between the QD and the cavity center. We estimate that the positioning accuracy is limited to 
$\pm150$ nm. We note that most of the measured points lie below the shaded region because the resolution limit 
of the OMA does not allow the true peak amplitude to be measured leading to an underestimate of the EF. 
 
To estimate the true EF, we assumed that $\Delta\lambda_{\rm true}=\Delta\lambda^y$, where $\Delta\lambda^y$ is the 
measured FWHM of the $y-$mode transmission peak which is non resolution-limited.
Using the measured values of $\Delta\lambda^y$, the true EF can be estimated using the formula 
 $E_{\rm true} = (E_{\rm meas.}-1)\Delta\lambda_{\rm OMA}/\Delta\lambda^y + 1$, where the factors of 1 account for the 
 normalized background  at $\lambda_{\rm res}$.
 The black circles in Fig.~\ref{fig:enhance}(c) show EFs corrected using the above formula. 
 We note that essentially all the points lie inside the shaded region within the experimental errors.

As seen in Fig.~\ref{fig:enhance}(c),
we observed EFs which coincided with the maximum value predicted by simulations at $\lambda_{\rm res}$ 
values of $785$ nm, $790$ nm, and $795$ nm. This suggests that for these cases, the QD position was close 
to one of the central cavity antinodes with $\phi$ close to zero. Taking as an example the case where $\lambda_{\rm res}=785$ nm,
the maximum corrected measured EF was $15\pm3$, in good correspondence with the maximum simulation value of $12$.

We note that the EF is given by $E = \gamma_g^{(c)}/3\langle\gamma_g^{(0)}\rangle$, where $\gamma_g^{(c)}$ is the decay rate into the guided modes of the CPCC,
and $\langle\gamma_g^{(0)}\rangle$ is the polarization average of the decay rate into the bare nanofiber guided modes.
The factor of $3$ arises because all the polarizations contribute to the background PL intensity but only one polarization contributes to the enhancement peak.
We can write $\gamma_g^{(c)} = (\gamma_g^{(c)}/\Gamma)(\Gamma/\Gamma_0)\Gamma_0$, where the first bracketed term is the channeling efficiency $\eta_c$,
the second bracketed term is the Purcell factor $F_{\rm P}$, and $\Gamma$ ($\Gamma_0$) is the total (free-space) decay rate of the QE.
Note that the EF is \emph{not} equal to the Purcell factor in general. From the measured EF of $15\pm3$ as given in the preceeding paragraph,
we expect the Purcell factor for a $y-$polarized dipole emitter $F_{\rm P}^y$ and the channeling efficiency 
into the nanofiber guided modes $\eta_c^y$ to be $7$ and $0.65$ respectively at $\lambda_{\rm res}=785$ nm from simulations.

On the other hand, in the Purcell regime, $F_{\rm P}$ is approximately equal to the cooperativity $C$~\cite{CharmichaelBook}. 
 Therefore, we have $F_P\approx C=(2g)^2/(\kappa\Gamma_0)$,
where $g$ is the QE-cavity coupling, and $\kappa$ is the cavity decay rate. It may be shown that for
nanofiber cavities, $g=\sqrt{\gamma_g^{(0)}/\tau_L}$, where $\tau_L$ 
is the cavity traversal time~\cite{LeKien}. Using the identity $\kappa = \pi/(F_c\tau_L)$, where $F_c$ 
is the cavity finesse, the Purcell factor can be expressed as 
$F_{\rm P} \approx (4/\pi)PF_c$, 
where $P = \gamma_g^{(0)}/\Gamma_0$.
Taking $P=0.2$ at $\lambda_{\rm res}=785$ nm, as calculated by simulations, and $F_c=28\pm 1$ as calculated from the experimentally measured $y-$mode FWHM,
the simulation value for $L_{\rm eff}$ and assuming a nanofiber effective refractive index of $1.19$, we find $F_{\rm P}^y=6.9\pm0.2$, in good agreement with the value of $7$ independently calculated in the preceding paragraph. 

Note that $F_P$ has a non
cavity-dependent component due to the transverse field confinement of the bare nanowaveguide as characterized by $P$, 
along with a cavity dependent component characterized by $F_c$. This suggests two routes to achieving better EFs in future CPCC designs: increase the duty cycle or 
slat number to enhance cavity reflectivity and thus $F_c$, or  
reduce the nanofiber diameter thereby increasing $P$. Simulations indicate that using the above strategies, along with a more rectangular slat shape we could achieve 
a maximum Purcell factor several times larger than the value found in the present study. Such a value would be competitive 
with PhC cavities in higher refractive index materials~\cite{Hausmann}.

The results presented here demonstrate that it is possible to achieve cavity based enhancement of SE for QEs placed on the 
surface of a nanofiber.
The CPCC method is flexible and can be applied to nanowaveguides other than nanofibers. 
We anticipate that the composite technique of realizing cavity-enhanced SE on a nanowaveguide as demonstrated here will provide a 
flexible new tool for future research in PhC cavity enhanced 
light-matter interactions and may open a new route to the realization of quantum networks.

This work was supported by the Japan Science and Technology Agency (JST) as one of the Strategic
Innovation projects.

%
%

{ \tiny
 }
\end{document}